# 大规模并行自适应有限元计算中的负载平衡研究

刘辉, 崔涛*, 冷伟, 张林波

Dynamic load balancing for large-scale adaptive finite element computation
Hui Liu, Tao Cui, Wei Leng, and Linbo Zhang

**摘要**

偏微分方程的并行求解, 关键问题之一是网格划分, 它不仅要求每个进程拥有相等的计算负载, 同时要求有良好的划分质量, 以减少进程间通信. 在自适应有限元计算过程中, 网格/基函数不断调整, 会导致负载不平衡, 必须动态地调整网格分布, 从而实现动态负载平衡. 本文研究了不同的负载平衡方法, 并在并行自适应有限元平台PHG中实现. 数值实验表明我们的动态负载平衡算法具有很高的划分质量, 运行速度快, 可有效划分网格并减少运行时间.

**关键字**: 自适应有限元, 并行计算, 动态负载平衡, 空间填充曲线, 加密树.

For the parallel computation of partial differential equations, one key is the grid partitioning. It requires that each process owns the same amount of computations, and also, the partitioning quality should be proper to reduce the communications among processes. When calculating the partial differential equations using adaptive finite element methods, the grid and the basis functions adjust in each iteration, which introduce load balancing issues. The grid should be redistributed dynamically. This paper studies dynamic load balancing algorithms and the implementation on the adaptive finite element platform PHG. The numerical experiments show that algorithms studied in this paper have good partitioning quality, and they are efficient.

## 1 引言

偏微分方程的并行数值求解的关键因素之一是处理好进程间的数据分布. 数据分布指根据进程数对数据进行划分, 将数据分布存储在各进程中. 计算过程中数据分布保持不变的程序, 如传统有限元程序, 在整个计算过程中仅需在开始做一次静态划分. 然而对自适应有限元程序, 每个进程的负载是在不断地变化的, 并且这个变化是不可以或很难预测的. 这种情况下, 就需要动态地调整数据分布以保持负载平衡. 好的数据划分方法既要使得每个进程拥有相同的计算负载, 又要减少进程间的通信.

并行自适应有限元计算中最常用的数据划分方法是划分计算网格, 每个进程负责一个子网格上的计算, 为保证可扩展性, 每个进程仅存储该子网格及定义在其上的数据, 不在本地的数据在需要时通过进程间通信获得. 这种计算模型包含三个模块: 数据划分、维护本地数据结构以方便计算和进程间通信、在需要的时候进行通信. 在自适应有限元程序中, 工作负载是与网格的单元、面、边及顶点相关的. 数据划分问题就是将这些对象划分到不同的进程中. 并行计算中的数据划分方法可以分为两类: 基于几何性质的划分方法和基于图的划分方法 [6, 7, 9, 10]. 基于几何性质的划分方法是利用对象 (粒子、单元等) 的几何信息划分数据, 如坐标, 常用的方法有递归坐标二分方法 RCB (Recursive Coordinate Bisection) [22, 23]、递归惯量二分方法 RIB (Recursive Inertial Bisection) [24]、空间填充曲线划分方法 SFC (Space-Filling Curve) [33, 12, 26] 等. 基于图的划分方法是利用对象的拓扑信息, 如网格中单元的邻居关系, 常用方法有递归图二分方法 (Recursive Graph Bisection)、贪婪算法、递归谱二分方法 RSB (Recursive Spectral Bisection) [24, 25]、K-L 算法 (Keinighnan-Lin Algorithm)、多水平方法 (multilevel methods) [18, 19, 14, 15, 17] 和扩散方法 (diffusive methods) [20, 21] 等. 图方法和几何方法各有优缺点. 图方法运行时间长, 由于显式地控制通信量, 划分质量高. 几何方法在空间局部性很重要或者拓扑结构不存在的情况下是非常有效的, 几何方法没有显式控制通信量, 它只是根据空间位置信息划分, 可能会导致通信量很大, 但几何划分方法简单, 它们容易实现并且速度快. 同时, 像递归坐标二分方法、空间填充曲线划分方法等, 它们是隐式增量的 [6], 数据迁移的代价比较小. 在并行自适应有限元计算中, 最终目标是极小化总

---

体计算时间, 包括划分时间、数据迁移时间及有限元求解时间. 它们是相关的: 质量高的划分会减少有限元计算时间, 但可能会需要更多的划分时间; 质量差的划分需要时间少, 但可能增加有限元计算时间. 在实际应用中, 需要根据实际情况综合考虑这些因素. 本文在PHG [1, 2, 34, 37] 中实现了加密树划分算法, 空间填充曲线算法和一维划分算法. 在数值实验中, 这些方法与ParMETIS 以及Zoltan进行了比较, 结果表明我们的算法速度快, 划分质量高, 可减少程序运行时间. 动态负载平衡对并行程序的开发是很关键的[1, 4], 基于拓扑的划分算法对线性方程组的开发也很重要, METIS 和ParMETIS 被广泛应用在解法器开发中[39, 5, 8, 11, 13, 16, 27, 30, 28]. 本文的方法已经应用在PHG, 同时也有一些开源软件提供本文提到的方法例如Zoltan 和ParMETIS [17].

本文的安排如下. 在第二节对划分方法进行了细致地讨论, 给出了加密树划分算法的细节以及空间填充曲线算法的细节, 并讨论了一维划分算法和子网格-进程映射算法. 接下来给出了数值实验, 最后给出了结论.

## 2 网格划分方法

PHG 提供了与软件包 ParMETIS[35] 和 Zoltan[38, 38] 的接口, 前者提供了多水平图划分方法, 后者提供了超图划分方法以及常用的几何划分方法, 如递归坐标二分方法、递归惯量二分方法、希尔伯特空间填充曲线划分方法. 除此之外, 我们在PHG中实现两种网格划分方法: 加密树划分方法和空间填充曲线划分方法 [33, 6, 7, 9]. 本节下面部分对这两个方法及实现进行介绍.

### 2.1 加密树划分方法

PHG 存储自适应过程中生成的加密树, 利用加密树划分方法实施网格划分是一个很自然的想法. 加密树划分方法 (refinement-tree partition method) 由 William Mitchell 提出, 该方法基于加密/放粗过程中产生的二叉树, 它调用深度优先遍历算法访问二叉树, 按照访问顺序对网格单元排序. 二叉树的遍历需要满足条件: 先访问左孩子, 然后是右孩子. 由于要求当前叶子结点与下一个被访问的叶子结点有一个共享面, 加密树划分方法有良好的划分质量.

在加密树中按下面方法给每一个结点赋予一个权重 $w$: 叶子结点的权重由用户给定, 非叶子结点的权重为以该结点为根的子树的所有叶子结点的权重和. Mitchell 的算法由两步组成, 第一步先计算每一个结点的权重, 第二步将加密树二分, 使得每个集合(set) 中的叶子数目相等, 递归调用二分算法来完成网格划分. 按照 Mitchell 的分析, 算法的计算复杂度为 $O(N \log p + p \log N)$, 其中 $N$ 为叶子结点数目.

算法需要计算每一个结点的权重, 由于父单元在多个进程中同时存在, 在计算父单元的权重时, 通信比较复杂. 我们重新设计了算法, 为每一个叶子结点定义前缀和 (prefix sum). 按照叶子结点的访问顺序, 叶子结点的前缀和定义为: 在到达该叶子结点之前访问的所有叶子结点的权重之和. 下面给出形式化定义, 按照叶子结点的访问顺序, 为每一个叶子结点赋予一个位置编号, 从 0 开始, 则第 i 个叶子结点的前缀和为

$$S_i = \sum_{j=0}^{i-1} w_j \tag{1}$$

其中 $w_j$ 为叶子结点 $j$ 的权重, 式 (1)可以改写为

$$S_i = \sum_{j=0}^{i-1} w_j = S_q + \sum_{j=q}^{i-1} w_j, \ (q < i). \tag{2}$$

假设所有叶子结点的权重之和为 $W$, 进程数为 $p$, 那么所有前缀和属于区间 [W i/p, W (i+1)/p) 的叶子单元分配给子网格 $i$ ($0 \leq i < p$). 通过分析可以看出, 只要计算出叶子结点的前缀和, 就可以通过计算叶子结点所属区间来完成划分.

当网格位于一个进程中时, 仅需遍历一次加密树即可完成划分, 算法复杂度为 $O(N)$. 当网格分布在不同进程中时, 假设进程数为 $p$, 每个进程含有的叶子结点数目分别为 $n_i$ ($0 \leq i < p$), 进程 $i$ 上叶子结点权重之和为 $W_i$ ($0 \leq i < p$). 定义叶子单元在本进程中的局部位置编号, 即叶子结点按照在本进程中访问顺序从 0 开始的编号. 定义 $S_{i,j}$ 为进程 i 上局部编号为 j 的叶子单元的前缀和, 那么由式 (2) 可以推导出

$$S_{i,j} = \sum_{k=0}^{i-1} W_k + \sum_{k=0}^{j-1} w_j = S_{i,j-1} + w_{j-1}, \tag{3}$$

上式表明, 只要知道了 $W_i$ ($0 \leq i < p$) 就可以在每一个进程中计算出所有叶子结点的前缀和. 在每个进程上, 遍历一次加密树就可以知道本进程所有叶子结点的权重和 $W_i$ ($0 \leq i < p$), 然后通过第二次遍历就可以同时完成计算前缀和及网格划分. 算法如下所示.

本文设计的算法简洁, 仅需遍历两次加密树以及一次 `MPI_Scan` 通信, 算法的计算复杂度为 $O(N)$. 对加密树划分算法, 由于初始网格会包含许多单元, 因而也会有多个加密树, 加密树的访问顺序是按照根结点的顺序, 只需

**Algorithm 1** PHG 实现的加密树划分算法

***Step 1.*** 每个进程访问本地的子树, 计算所有本地叶子结点的权重和 $W_i$.

***Step 2.*** 调用 MPI_Scan 操作, 为每个进程收集其需要的 $W_i$ $(0 \le i < p)$.

***Step 3.*** 访问本地子树, 根据式 (3) 计算每一个叶子结点的前缀和, 并同时完成划分.

对初始网格 (根结点) 排序, 并在整个自适应过程中维护这个序, 以保证在整个自适应过程中访问子树的顺序是相同的.

## 2.2 空间填充曲线划分方法

网格划分中利用空间填充曲线将高维空间映射成一维空间, 完成单元的排序, 从而将高维划分问题转化为一维划分问题. Morton 空间填充曲线和希尔伯特空间填充曲线均具有较好的空间局部性, 划分质量比较好, 是最常用的两种空间填充曲线 [33, 12, 26].

空间填充曲线划分方法分为三个步骤. 第一个步骤是计算空间填充曲线, 首先将计算区域映射到立方体 $(0,1)^3$, 然后调用空间填充曲线生成算法将立方体映射到区间 $(0, 1)$; 第二个步骤是划分区间 $(0, 1)$, 使得属于每个区间单元的权重和相等; 第三个步骤是调用子网格–进程映射算法, 对子网格重新编号以减少数据迁移.

对计算区域 $\Omega$, 存在一个长方体 (三维, 在二维情况下为长方形) 包围盒 (bounding box) 包含这个区域, 假设包围盒在 $x, y, z$ 三个方向的长度分别为 $\text{len}_x, \text{len}_y, \text{len}_z$, 将区域映射到 $(0,1)^3$ 的通常的做法是

$$x_1 = (x - x_0) \,/\, \text{len}_x, \ y_1 = (y - y_0) \,/\, \text{len}_y, \ z_1 = (z - z_0) \,/\, \text{len}_z,$$

其中 $(x_0, y_0, z_0)$ 是包围盒坐标最小的顶点, $(x_1, y_1, z_1)$ 为坐标 $(x, y, z)$ 经过变换后在立方体 $(0,1)^3$ 中的坐标. 上述变换改变了区域的长宽比, 使得变换后的长宽比为 $1:1:1$, 破坏了区域的空间局部性. PHG 中取 $\text{len} = \max(\text{len}_x, \text{len}_y, \text{len}_z)$, 采用变换

$$x_1 = (x - x_0) \,/\, \text{len}, \ y_1 = (y - y_0) \,/\, \text{len}, \ z_1 = (z - z_0) \,/\, \text{len},$$

该变换保持区域的空间局部性, 改善了划分质量, 这一点将在后面用数值算例加以验证.

对空间填充曲线划分算法, 我们提供了两种空间填充曲线生成程序: Morton 空间填充曲线和希尔伯特空间填充曲线. 前者的算法简单, 但曲线本身有较大的跳跃, 使得空间局部性略差. 希尔伯特空间填充曲线的空间局部性要好很多, 但生成算法复杂.

首先利用空间填充曲线对单元排序, 然后调用 2.3 中介绍的一维划分算法及 2.4中介绍的子网格–进程映射算法, 便可完成网格划分.

## 2.3 一维划分算法

上节的空间填充曲线划分方法及其他一些网格划分方法最终均将问题转化为一维划分问题, 因此实现一个高效健壮的一维划分程序是必要的.

假设当前进程数为 $p$ 以及给定区间 [a, b), 这里的一维划分问题为: 如何将 [a, b) 划分成 $p$ 个子区间 $[a, a_1)$, $[a_1, a_2), \cdots, [a_{p-2}, a_{p-1}), [a_{p-1}, b)$, 使得每个区间上包含的对象的权重和相等.

从问题的描述可以看出, 一维划分需要计算 $p - 1$ 个分割点 $a_1, a_2, \cdots, a_{p-2}, a_{p-1}$. 我们在 PHG中实现了一个一维划分算法求解该问题, 其基本算法来源于 Zoltan[38], 它是二分搜索算法的推广. 在二分搜索算法中, 只需求解一个点 $a_1$, 将区间分成两个子区间. 如果将区间 $k$ $(k \ge 2)$ 分而不是二分, 那么 $a_1$ 将位于某个子区间中, 接着将这个子区间 $k$ 分, 重新得到 [a, b) 的一个划分, $a_1$ 将位于上述 $k$ 个更小的子区间中的一个. 重复这个过程, 可以在给定的精度下搜索到 $a_1$. 将这个算法推广, 将区间 [a, b) 划分成 $N$ $(N \gg p)$ 个子区间, $a_i$ $(0 < i < p)$ 将位于某个子区间, 然后将子区间细分, 最终在给定的精度下可以搜索到 $a_1, a_2, \cdots, a_{p-2}, a_{p-1}$.

具体实现时, 我们取子区间数为 $N = (p-1)*k+1$, 程序为每一个划分点 $a_i$ $(0 < i < p)$ 维护一个包围盒 (bounding box). 在每一次划分之前先更新这些包围盒, 缩小 $a_i$ $(0 < i < p)$ 的取值范围, 重新划分区间 $[0, 1)$ 的时候, 直接划分 $a_i$所在的包围盒, 而不是 $a_i$ $(0 < i < p)$ 所在的某个子区间, 这样会加速求解 $a_i$ $(0 < i < p)$ 的过程.

## 2.4 子网格－进程映射

网格划分完成后, 需要将子网格映射到进程上, 该映射尽量使得从老划分到新划分的数据迁移量最小. Oliker 和 Biswas 设计了一个启发式算法处理这个问题, 他们的算法可以得到次优解 [29]. 我们在 PHG中实现了该算法.

Oliker 和 Biswas 的算法先对数据迁移建模. 用相似矩阵 S (similarity matrix) 表示数据在所有进程中的分布情况. S 在 PHG 中是一个 $p_{\text{old}} \times p$ 的矩阵, $p_{\text{old}}$ 是当前子网格数, $p$ 为新子网格数 (在通信器不变的情况下, $p_{\text{old}}$ 与 $p$ 是相等的), 相似矩阵 S 的元素 $S_{i,j}$ 表示编号为 $i$ 的子网格中需要迁移到编号为 $j$ 的子网格中的数据量. 实际计算中, 每个进程并发地计算相似矩阵的一行, 然后通过一个主进程收集这些信息创建一个相似矩阵, 根据这个相似矩阵计算子网格–进程映射关系, 最后通过一个广播操作将映射关系发送给所有进程.

优化子网格–进程映射时, 需要针对不同的体系结构上需要建立不同的代价方程 (cost function) 以优化数据迁移开销. 常用的两种度量是: `TotalV` 和 `MaxV`. `TotalV` 极小化所有进程数据迁移量之和; `MaxV` 极小化单个进程的最大数据迁移量. Oliker 和 Biswas 的算法中使用的度量是 `TotalV`. 给定一个相似矩阵 S, 其数据总和是确定的, 极小化数据迁移与极大化保持数据不动是等价的, 目标是寻找一个映射 $i \rightarrow p_i$ $(i = 0, \cdots, p-1)$ 极大化如下代价方程

$$F = \sum_{\substack{p_j = i \\ 0 \leq j < p}} S_{i,j},$$

其中 $(p_0, p_1, \cdots, p_{p-1})$ 是 $(0, 1, \cdots, p-1)$ 的一个置换.

## 3 数值算例

本节给出数值算例, 比较不同划分方法的性能、划分质量以及对有限元计算性能影响. 数值算例使用自适应有限元方法求解偏微分方程, 比较不同划分方法对程序性能的影响. 网格划分算法只负责划分, 数据迁移在PHG 中是一个独立的模块, 用来完成网格, 自由度数据等的重建.

本节使用了网格 $\Omega_1$ 是一个圆柱体, 如图 3.1 所示, 直径小但长度较长, 有较大的长宽比, 网格的单元数目为 2,522,624. 网格由 `Netgen`[32] 生成.

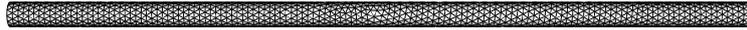

图 3.1: 测试网格 $\Omega_1$ (圆柱体)

### 3.1 划分质量对自适应有限元计算性能的影响

本节将 ParMETIS, Zoltan/HSFC, PHG/HSFC, PHG/RTK (PHG 的加密树划分方法), MSFC (PHG中的 Morton 空间填充曲线划分方法), RCB (Zoltan的递归坐标二分方法) 六种划分方法应用到自适应有限元计算中, 以研究不同划分方法对并行程序总体性能的影响.

测试的硬件平台为科学与工程计算国家重点实验室的浪潮 TS10000 高性能集群 (LSSC-III), 计算结点为浪潮 NX7140N 刀片, 每个刀片包含两颗 Intel X5550四核处理器和 24GB 内存, 其单核双精度浮点峰值性能为 10.68Gflops, 282 个计算结点的总浮点峰值性能为 24Tflops. 所有结点同时通过千兆以太网和 DDR Infiniband 网络互联.

**算例3.1.** 本例求解如下 `Dirichlet` 边界条件 `Helmholtz` 方程,

$$\begin{cases} -\Delta u + u = f & (x,y,z) \in \Omega \\ u(x,y,z) = g & \text{on } \partial\Omega. \end{cases}$$

计算区域为圆柱体区域 $\Omega_1$, 取解析解如下:

$$u = \cos(2\pi x)\cos(2\pi y)\cos(2\pi z),$$

该算例中解是光滑的, 因此网格加密基本上是均匀的.

测试中使用了 32 个计算结点, 128 个进程. 计算区域如图 3.1所示, 初始网格 $\mathcal{T}_1$ 由 Netgen [32] 生成, 含有 4,927 个四面体单元. 使用三阶协调拉格朗日元对方程进行离散, 调用数值代数软件包 `Hypre` 的 `BoomerAMG` 解法器求解有限元离散产生的线性系统.

图 3.2 给出划分时间, 从中可以看出 RTK 方法的划分速度最快, 其次是 MSFC, PHG/HSFC, Zoltan/HSFC. PHG实现的 HSFC 及 MSFC 均比 Zoltan/HSFC 速度要快. ParMETIS 和 RCB 最慢. 从图上可以看出网格分

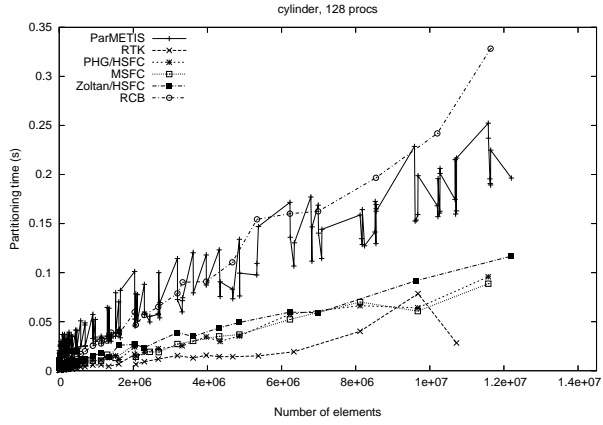

图 3.2: 网格划分时间 (例 3.1)

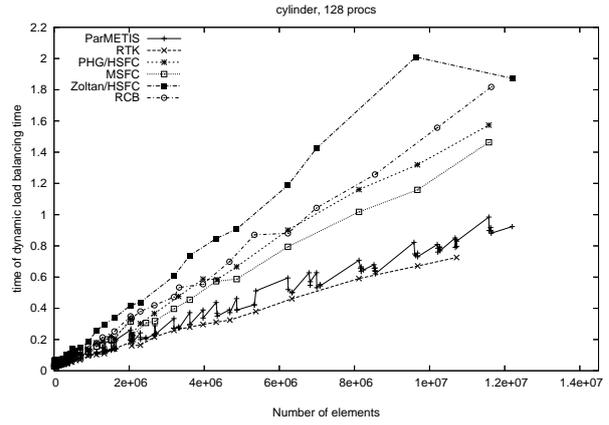

图 3.3: 动态负载平衡时间 (例 3.1)

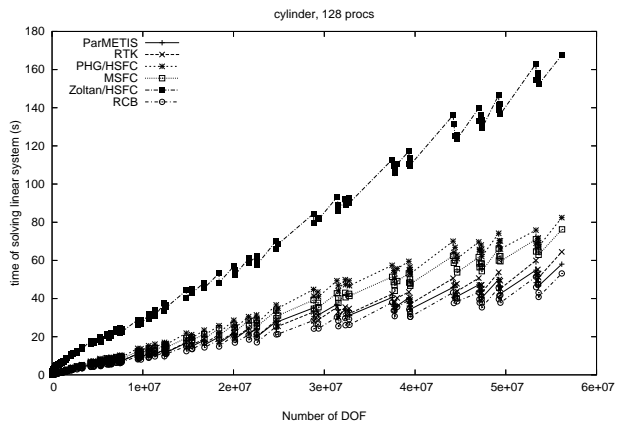

图 3.4: 求解时间, 横坐标为自由度数 (例 3.1)

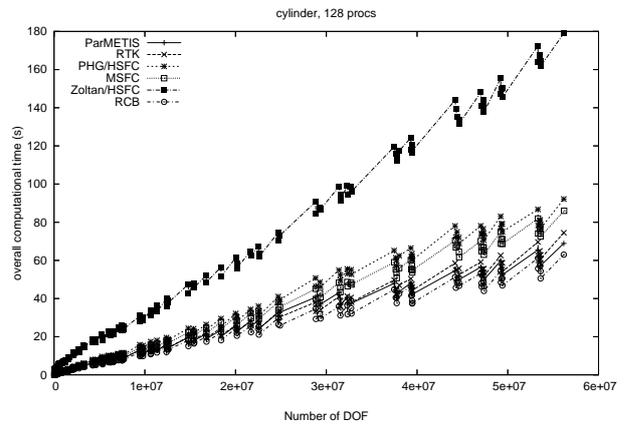

图 3.5: 每一个自适应步时间 (例 3.1)

布对 ParMETIS 的划分时间有较大影响, 随着网格调整, 其划分时间出现剧烈震荡. 总体来讲, 几何划分方法对网格分布不敏感, RCB, MSFC, PHG/HSFC, Zoltan/HSFC 的划分时间随着网格规模的增大以较稳定的速率增加.

图 3.3 给出动态负载平衡时间, 其中包含了网格划分和数据迁移时间, 其中数据迁移时间占的比重较大. 一个好的划分方法应该是增量的, 即网格的微小变化仅导致划分的微小变化, 这样会使得数据迁移量比较小. 从图上可以看出 RTK 的时间最短, 表明其数据迁移量较少, RTK 的时间增长也很稳定. 其次是 ParMETIS, ParMETIS同样表现出震荡. 接着是 MSFC, PHG/HSFC, RCB, 用时最长的是 Zoltan/HSFC. 它们的时间增长趋势均比较稳定, 说明这些方法均有较好的可扩展性和比较低的计算复杂度.

一般情况下线性方程组的求解时间在自适应有限元计算中占统治地位, 图 3.4 和 3.5 分别给出线性方程组求解时间和每一个自适应步的时间. 从图中可以看出, RCB, ParMETIS, RTK 运行时间最短. 对 RCB 而言, 其划分质量一般要比 ParMETIS 及 RTK 差一些, 这个算例是特例, 因为计算区域是一个长圆柱体, 非常适合 RCB. 其次是 MSFC, PHG/HSFC, 用时最长的是 Zoltan/HSFC. Zoltan/HSFC 与 PHG/HSFC 是同一类型的划分方法, 由于 PHG在映射时采取了保持区域长宽比的方式, 维持了区域的空间局部性, PHG/HSFC 的划分质量比 Zoltan/HSFC 高, 求解时间也较少.

Table 1: 总体运行时间及划分次数 (例 3.1)

| Method | total running time(s) | # of repartitionings |
| --- | --- | --- |
| RCB | 3049.60 | 60 |
| ParMETIS | 3360.73 | 189 |
| RTK | 3465.63 | 59 |
| MSFC | 4088.01 | 58 |
| PHG/HSFC | 4493.43 | 59 |
| Zoltan/HSFC | 8954.21 | 50 |

表 1 给出该算例采用不同划分方法时的运行时间及划分次数, 算例包含 190 个自适应步. RCB 运行时间最短, 说明对计算区域比较规则的问题, 简单的几何划分方法是非常有效的. Zoltan/HSFC 的运行时间最长, 是别的方法的两倍以上, 表现最差. ParMETIS 的划分次数最多, 是别的方法的三倍. 划分次数过多会导致数据迁移频繁, 当网络带宽较低时, 就需要考虑数据迁移问题, 但由于其划分质量好, 线性方程组的求解时间较短.

**算例3.2.** 本例求解的方程为线性抛物型方程

$$\begin{cases} u_t - \Delta u = f & \text{in } \Omega \times (0, T) \\ u = g & \text{on } \Gamma \times [0, T]. \end{cases} \tag{4}$$

取解析解为

$$u = \exp((25((x - \frac{1}{2} - \frac{2}{5}\sin(8\pi t))^2 + (y - \frac{1}{2} - \frac{2}{5}\cos(8\pi t))^2 + (z-1)^2) + 0.9)^{-1} - 2.5).$$

计算区域为 $\Omega_3 = (0,1)^3$, 时间区间为 $[0,1]$. 解的极大值在 $z=1$ 平面上运动, 网格在极值附近最密. 由于这是一个依赖于时间的问题, 网格自适应同时包含加密和放粗.

Table 2: 程序运行时间 (例 3.2, 128 进程)

| Method | Time TAL(s) | Time DLB(s) | Time SOL (s) | Time STP(s) |
|---|---|---|---|---|
| PHG/HSFC | 6525 | 0.0734 | 0.1886 | 0.9192 |
| Zoltan/HSFC | 6744 | 0.0917 | 0.1928 | 0.9501 |
| MSFC | 6902 | 0.0730 | 0.1966 | 0.9724 |
| RTK | 7015 | 0.0738 | 0.2050 | 0.9884 |
| RCB | 7131 | 0.1126 | 0.1938 | 1.0046 |
| ParMETIS | 7151 | 0.1421 | 0.2114 | 1.0075 |

Table 3: 程序运行时间 (例 3.2, 192 进程)

| Method | Time TAL(s) | Time DLB(s) | Time SOL(s) | Time STP(s) |
|---|---|---|---|---|
| PHG/HSFC | 6597 | 0.0932 | 0.1808 | 0.9294 |
| MSFC | 6601 | 0.0936 | 0.1863 | 0.9299 |
| Zoltan/HSFC | 6646 | 0.1046 | 0.1862 | 0.9362 |
| RCB | 7197 | 0.1176 | 0.1862 | 1.0139 |
| RTK | 7185 | 0.0799 | 0.2124 | 1.0123 |
| ParMETIS | 7218 | 0.1982 | 0.1942 | 1.0169 |

这里给出两组测试结果. 第一组测试使用 32 个计算结点、128 个进程, 共运行了 7098 个时间步, 每个时间步的平均单元数量为 663,151, 平均自由度数为 919,036. 结果见表 2, 其中第一项时间为总运行时间 (TAL), 第二项为单次动态负载平衡的平均时间 (DLB), 第三项为单次线性方程组的平均求解时间 (SOL), 第四项为单个时间步的平均计算时间 (STP). 从表 2 中可以看出, 几何划分方法的效果优于图划分方法, PHG/HSFC, Zoltan/HSFC, 及 MSFC 均优于 RTK 及 ParMETIS, RCB 除外. 其中 PHG/HSFC 运行时间最短, ParMETIS 运行时间最长. 可以看出, 当网格变化较剧烈时, 几何划分方法也有优势. 一般来说, 在静态划分中, 图划分方法划分质量最好, 最有优势; 在动态划分中, 情况则不一定.

第二组测试数据如表 3 所示, 其中使用了 32 结点、192 进程. 结果与第一组测试类似, 几何划分方法用时较少, 仍优于图划分方法. 在本算例的两个表格中, PHG/HSFC 均比 Zoltan/HSFC 用时略少, 说明 PHG 实现的版本的划分质量略高, 但没有前面两个算例的差别那么大, 因为在这个算例中, 计算区域为 $(0,1)^3$, Zoltan/HSFC 的映射结果和 PHG/HSFC 是类似的. 两个版本的差别在区域的长宽比较大时体现得更为明显.

# 4 本文小结

本文介绍了网格划分方法, 并通过数值实验比较了各种网格划分方法在并行自适应有限元计算中的优劣. 通常来讲, 几何划分方法的速度快、实现简单, 但划分质量不高; 图划分方法的速度慢、算法复杂, 但划分质量非常高. 对静态问题或者网格变化较小的问题, 图划分方法是最好的选择; 对网格频繁调整的问题, 几何划分方法是很有竞争力的. 在实际应用中, 应综合考虑网格划分、数据迁移、问题求解等多个因素确定合适的划分方法.